\pdfoutput=1
\documentclass{article}

\usepackage{graphicx}
\usepackage{subfigure}

\usepackage{amsmath}
\usepackage{amssymb}
\usepackage{amsfonts}
\usepackage[hidelinks]{hyperref}

\usepackage[affil-it]{authblk}

\usepackage{xcolor}
 
\newcommand{\citep}{\cite}
\newcommand{\email}[1]{\href{mailto:#1}{\nolinkurl{#1}}}

\newcommand\blfootnote[1]{%
  \begingroup
  \renewcommand\thefootnote{}\footnote{#1}%
  \addtocounter{footnote}{-1}%
  \endgroup
}

\begin{document}

\title{Stochastic Blockmodels with Edge Information}

\author[1]{Guy W. Cole}
\author[2]{Sinead A. Williamson\footnote{Contributions made prior to Amazon employment.}}

\affil[1,2]{The University of Texas at Austin, 2317 Speedway G2500, Austin, TX 78712}
\affil[2]{Amazon, Inc.}
\affil[1]{\email{guywcole@utexas.edu}}
\affil[2]{\email{Sinead.Williamson@mccombs.utexas.edu}}

\maketitle

\begin{abstract}
Stochastic blockmodels allow us to represent networks in terms of a latent community structure, often yielding intuitions about the underlying social structure. Typically, this structure is inferred based only on a binary network representing the presence or absence of interactions between nodes, which limits the amount of information that can be extracted from the data. In practice, many interaction networks contain much more information about the relationship between two nodes. For example, in an email network, the volume of communication between two users and the content of that communication can give us information about both the strength and the nature of their relationship.

In this paper, we propose the Topic Blockmodel, a stochastic blockmodel that uses a count-based topic model to capture the interaction modalities within and between latent communities. By explicitly incorporating information sent between nodes in our network representation, we are able to address questions of interest in real-world situations, such as predicting recipients for an email message or inferring the content of an unopened email. Further, by considering topics associated with a pair of communities, we are better able to interpret the nature of each community and the manner in which it interacts with other communities.  

\blfootnote{Author's note: This paper was developed concurrently and independently to Bouveyron et al. in \cite{bouveyron2018stochastic}, who develop a similar model, propose a different inference strategy, and apply it to the Enron data set as well as others.}

\end{abstract}
\section{Introduction}\label{sec:introduction}

A key focus in statistical network analysis has been the search for low-dimensional representations of the observed structure. One of the most commonly used frameworks is the stochastic blockmodel \citep{Wang:Wong:1987,Snijders:Nowicki:1997}, where nodes are assumed to belong to one of $K$ latent communities. 

Typically, the networks modeled using stochastic blockmodels are binary, and interactions are modeled as Bernoulli random variables. However, binary interaction networks contain minimal information about the relationship between each pair of nodes, leading to a weakly informative likelihood. The presence or absence of an interaction between two nodes conveys only a single bit of information, meaning that for moderately-sized networks the posterior distribution can be very disperse. This in turn makes it difficult to infer fine-grained structure.

Fortunately, in real-life social networks, we typically have more information about the interaction between two entities. For example, in an email network, the number of emails sent between two users can be seen as a proxy for interaction strength. Further, the content of emails may be used to offer more information regarding the nature of the relationship between two individuals. Despite this rich trove of information associated with interactions, there has been little attempt in the blockmodel literature to exploit text sent across a network in learning community structure.

We propose the Topic Blockmodel, a network model that represents the interaction between two nodes not as a binary indicator variable, but as the totality of their communication. 
Concretely, we assume that an interaction comprises a sequence of words, such as an email chain or a conversation. 
Each pair  of communities is associated with a count-based topic model  which governs both the volume and the content of interactions between members of those communities.

The benefits of this richer formulation are two-fold. First, by associating a pair of communities with a distribution over topics rather than just a probability of interaction, we improve interpretability of the  communities found. By considering the topics afforded high probability for a given community-community pair, we can automatically generate an interpretable label characterizing the pair.

Secondly, we can use the resulting model to ask questions of interest about the network. For example, an email provider might wish to suggest recipients for an email being composed. By considering both the set of people with whom the author has previously corresponded and the text of the composed email, the Topic Blockmodel can make better predictions than a binary or integer-valued stochastic blockmodel. Another example might be flagging emails in a security application, where we want to identify emails on a given topic: by jointly modeling interactions and topics, we can use community information to make predictions about the topical content of an email based on its sender and recipient, even if the email is encrypted.

We begin in Section~\ref{sec:background} by reviewing the stochastic blockmodel framework, and discussing existing methods that incorporate both network and topic information. We then present the Topic Blockmodel in Section~\ref{sec:model}. After briefly describing inference in Section~\ref{sec:inference}, we showcase the performance
of the Topic Blockmodel on real data in Section~\ref{sec:results}. By looking at a naturally generated network of emails, and a semi-realistic network based on characters in a play, we demonstrate that the Topic Blockmodel yields both interpretable clusters, and impressive predictive performance  both in terms of recipient prediction given a communication's text and author, and topic prediction given a communication's sender and recipient. Finally, we discuss possible extensions in Section~\ref{sec:conclusion}.

\section{Background}\label{sec:background}

The Topic Blockmodel presented in this paper is a stochastic blockmodel that incorporates both a count model and a topic model in its likelihood. In this section, we review both stochastic blockmodels and topic models, and discuss existing models that combine these approaches.

\subsection{Stochastic Blockmodels}

Stochastic blockmodels \citep{Wang:Wong:1987,Snijders:Nowicki:1997} are a popular class of generative models that assume that each node within a network is associated with one of $K$ latent clusters or communities. Each pair $(k,\ell)$ of communities is associated with a latent parameter $\lambda_{k,\ell}$, which parametrizes the interactions between members of those communities. Typically, the network is assumed to be binary, and the interactions are modeled as Bernoulli random variables. In a Bayesian setting, we place conjugate (in the binary case, beta) priors on the $\lambda_{k,\ell}$, and a Dirichlet-multinomial prior on the community memberships.

A number of variants to the basic stochastic blockmodel have been proposed. \citep{Karrer:Newman:2011} uses a gamma/Poisson link in place of a beta/Bernoulli, to obtain distributions over integer-valued networks, and also incorporates a per-node parameter that allows nodes in the same community to have different degree distribution. The Infinite Relational Model \citep{irm} allows a potentially infinite number of communities, with membership probabilities distributed according to a Dirichlet process. Rather than restrict each node to a single cluster, the Mixed Membership Stochastic Blockmodel \citep{Airoldi:Blei:Fienberg:Xing:2008} associates each node with a distribution over clusters, allowing nodes to perform several social roles.

In this work, we adopt the Poisson links introduced by \citep{Karrer:Newman:2011} to capture the communication volume between nodes. Our model could be extended to incorporate the nonparametric and mixed membership behavior described above; however as we discuss in Section~\ref{sec:conclusion}, this would significantly increases the computational cost of the model and we leave this for future work.

\subsection{Topic Models}
Topic models are a popular family of hierarchical Bayesian models for semantic analysis of corpora of documents.  The canonical model of this type is Latent Dirichlet Allocation \cite{blei2003LDA}, where each document is associated with a Dirichlet-distributed distribution over $T$ ``topics'', which themselves are Dirichlet-distributed distributions over words that tend to concentrate on semantically coherent topics. Each word  in the  document is assumed to have been generated by sampling a topic from that document's distribution over topics, and then sampling a word from the topic's distribution over words. 

This basic model has been extended in a number of directions. A hierarchy of Dirichlet processes can be used to construct a Bayesian nonparametric variant with an unbounded number of topics \citep{Teh:2007}; a logistic normal distribution can be used to induce correlations between topics \citep{ctm}; time dependence has been incorporated to track topic evolution over time \citep{blei2006dynamic}. While this work builds upon the basic  LDA topic model, incorporation of more complex topic models is an interesting avenue for future exploration.

\subsection{Existing Network-Based Topic Models}\label{sec:related}
A number of works have attempted to combine network and topic models. 
They loosely fall into two camps: models that treat the network as a fixed covariate used to guide the topic model; and models that jointly model a corpus of documents and an associated network. An example of the first type of model is the Author-Recipient Topic Model \citep{mccallum2005}, which uses the network to specify a separate  topic distribution for each sender-recipient pair. This does not allow for the elucidation of community structure, or provide conditional distributions over recipients.

The second type of model treats the network and the text as two related datasets described using a single probabilistic model.
The Relational Topic Model \citep{rtm} and the Poisson mixed-topic link model \citep{Zhu:Yan:Getoor:Moore:2013} use the topic assignments of two documents to determine the probability of an interaction between them, resulting in a binomially distributed number of links associated with each document. The Citation Author Topic Model \citep{tu2010} associates each topic with a distribution over words and a distribution over citable authors, and uses this to generate a set of interactions.

Another model that falls under this framework is the Joint Gamma Process Poisson Factorization (J-GPPF) model \citep{acharya2015}, which models interactions between nodes using an infinite blockmodel and associates each community with a distribution over topics; the topics associated with an author's community membership are used to generate documents written by that author. The J-GPPF model is the closest approach to our own, since it explicitly clusters users into communities and uses those communities to guide a topic model. However, like all the models described above, the J-GPPF model assumes the (binary) network is modeled as a distinct entity from the documents. This is appropriate where an individual is associated both with a collection of documents and a set of connections---for example, in a scientific setting, the documents might be an author's papers, and the connections might be the set of people they have cited. J-GPPF does not translate into our setting, where the network is implicitly defined by the text sent across it.

By contrast, rather than conditioning on the network, or modeling it jointly, our Topic Blockmodel explicitly uses a topic model as a link function in a stochastic block model.  Treating the text and the relationship as equivalent captures the idea that the collection of documents sent from node $s$ to node $r$ encapsulates their relationship. In this setting, we extract information not just from the fact that Alice sent emails to Bob about football; we also make use of the fact that Alice sent \textit{no} messages to Claire. This absence of a link between Alice and Claire is informative about the underlying community structure.

\section{Stochastic Blockmodel with Topic Links}\label{sec:model}

Following the basic stochastic blockmodel framework, we assume a distribution $\phi\sim \mbox{Dirichlet}_K(\xi_0)$ over $K$ communities, and associate each node $s$ with a cluster $c_s \sim\mbox{Discrete}(\phi)$ sampled from this distribution. We then associate each pair $(k,\ell)$ of communities with a set of parameters $\lambda_{k,\ell}$.

Unlike the binary stochastic blockmodel, where $\lambda_{k,\ell}\in (0,1)$ is a beta-distributed random variable used to parametrize Bernoulli links, we let $\lambda_{k,\ell}=(\lambda_{k,\ell}^{(1)},\dots  \lambda_{k,\ell}^{(T)})$ be a vector of gamma-distributed random variables. The $t$th element of this vector, $\lambda_{k,\ell}^{(t)}$, controls the number of words in topic $t$ that are sent from a member of community $k$, to a member of community $\ell$. Concretely, we let $n_{s,r}^{(t)}$, the number of words in topic $t$ sent from node $s$ to node $r$, be distributed according to $\mbox{Poisson}(\lambda_{c_s,c_r}^{(t)})$. The total number of words, $n_{s,r}^{(\cdot)} = \sum_{t=1}^Tn_{s,r}^{(t)}$, sent from node $s$ to node $r$ is therefore Poisson-distributed with parameter $\lambda_{c_s,c_r}^{(\cdot)}=\sum_{t=1}^T \lambda_{c_s,c_r}^{(t)}$. Marginally, $\lambda_{c_s, c_r}^{(\cdot)}$ is a $\mbox{Gamma}(T\alpha_\lambda, \beta_\lambda)$ random variable.

In order to complete our model specification, we specify a topic-specific distribution $\eta_t \sim \mbox{Dirichlet}_V(\kappa)$ over the size-$V$ dictionary for each of the $T$ topics. For each of the $n_{s,r}^{(t)}$ words associated with topic $t$, we then sample a word token according to $\eta_t$. The full generative process can therefore be summarized as
\begin{equation}
  \begin{aligned}
    \eta_t \sim& \mbox{Dirichlet}_V(\kappa),\; t\in \{1,\dots,T\}\\
    \phi \sim& \mbox{Dirichlet}_K(\xi_0)\\
    c_s \sim& \mbox{Discrete}(\phi),\; s\in\{1,\dots,S\}\\
    \lambda^{(t)}_{k,\ell}\sim& \mbox{Gamma}(\alpha_\lambda, \beta_\lambda),\; k,\ell \in\{1,\dots, K\}\\
    n_{s,r}^{(t)}\sim& \mbox{Poisson}(\lambda_{c_s,c_r}^{(t)})\; s,r \in\{1,\dots, S\}\\
    w_{s,r,i}^{(t)} \sim& \mbox{Discrete}(\eta_t),\; i\in\{1,\dots, n_{s,r}^{(t)}\}.
  \end{aligned}\label{eqn:PoisForm}
\end{equation}
Here, $w_{s,r,i}^{(t)}$ is the identity of the $i$th token sent from node $s$ to node $r$ under topic $t$. Rather than simply have two nodes' community memberships determine the probability of an interaction between them, in the Topic Blockmodel each pair of communities provides a distribution over the number of words sent in each of $T$ topics, determining both the overall volume of communication and its semantic content.

\sloppy An equivalent specification can be obtained by noting that, conditioned on the total number of words $n_{s,r}^{(\cdot)}$ sent from node $s$ to node $r$, the assignment of words to topics is given by a multinomial distribution parametrized by
$ \theta_{c_s,c_r} = \left(\lambda_{c_s,c_r}^{(1)},\dots,  \lambda_{c_s,c_r}^{(T)}\right)/\lambda_{c_s,c_r}^{(\cdot)}$.
Further, this vector of probabilities $\theta_{c_s,c_r}$ is independent of the normalizing constant $\lambda_{c_s,c_r}^{(\cdot)}$, and is distributed $\mbox{Dirichlet}_K(\alpha_\lambda)$. If we let $z_{s,r,i} =t$ if the $i$th word sent from node $s$ to node $r$ is in topic $t$, we can rewrite our model as
\begin{equation}
  \begin{aligned}
    \lambda^{(\cdot)}_{k,\ell}\sim& \mbox{Gamma}(T\alpha_\lambda),\; k,\ell\in\{1,\dots, K\}\\
    \theta_{k,\ell} \sim& \mbox{Dirichlet}_T(\alpha_\lambda)\\
    n_{s,r}^{(\cdot)}\sim& \mbox{Poisson}(\lambda_{c_s,c_r}^{(\cdot)})\; s,r\in\{1,\dots, S\}\\
    z_{s,r,i} \sim& \mbox{Discrete}(\theta_{c_s,c_r}),\; i\in\{1,\dots, n_{s,r}^{(\cdot)}\}\\
    w_{s,r,i} \sim& \mbox{Discrete}(\eta_{z_{s,r,i}}),
  \end{aligned}\label{eqn:DirMultForm}
\end{equation}
where the distributions over $\eta_t$, $\phi$ and $c_s$ are as given in Equation~\ref{eqn:PoisForm}.

These two equivalent formulations prove useful for inference. As we will see in Section~\ref{sec:inference}, the Dirichlet-multinomial formulation of Equation~\ref{eqn:DirMultForm} allows us to use standard LDA updates for the $z_{s,r,i}$. Conversely, the gamma-Poisson formulation of Equation~\ref{eqn:PoisForm} yields a straightforward-to-calculate likelihood for Gibbs sampling the cluster assignments. 

The Topic Blockmodel described above offers clear advantages over the models described in Section~\ref{sec:related}, without adding unnecessary complexity. In the models discussed previously, either the network was treated simply as a covariate, or it was modeled separately in a manner that assumes a marginally Binomial distribution over the number of recipients. Our model is appropriate in the setting where the documents \textit{are} the network, and the strength of an interaction is directly implied by the length of a document. In addition, we obtain latent community structure, which was not available from most of the models discussed previously.
\section{Inference}\label{sec:inference}

Since the hierarchical model is composed of conjugate pairs and we can separate the distribution over the total number of words from the conditional distribution over the nature of those words, construction of a Gibbs sampler is straightforward. Our sampler iteratively updates the community assignments $c_s$ for each node $s$, and the topic assignments $z_{s,r,i}$ for each word.

Conditioned on the community memberships $c_s$ and the number $n_{s,r}^{(\cdot)}$ of words sent from node $s$ to node $r$,  the updates for the topic assignments $z_{s,r,i}$ are standard LDA updates (see for example \citep{griffiths2004finding}), except with a topic mixture for each cluster pair rather than each document. 

Conditioned on the topic assignments, we can sample the cluster memberships according to
\begin{equation}
\begin{aligned}[l r] P(c_s=k|\mbox{rest}) & \propto (m_k^{-s}+\xi_0) \\ & \times \prod_{j=1}^K \prod_{t=1}^T P(\{n_{s,r}^{(t)}: c_r=j\}|c_s=k, \mbox{rest}) \\ & \times \prod_{j=1}^K \prod_{t=1}^T  P(\{n_{r,s}^{(t)}:c_r=j\}|c_s=k, \mbox{rest}), \end{aligned}
\end{equation}
where $m_k^{-s}$ is the number of nodes in community $k$ (excluding the $s$th node). The likelihood terms in the second and third line are straightforward to calculate due to gamma-Poisson conjugacy. 

\section{Experimental Evaluation}\label{sec:results}

In order to assess the interpretability and predictive power of the posterior obtained using the Topic Blockmodel, we ran experiments on two real-world datasets, comparing against a range of competing models. 

\subsection{Datasets}\label{sec:datasets}

We considered two datasets: A real-world email network and a network of fictional characters.

\paragraph{ENRON emails:}
The ENRON email dataset \citep{enron_paper} is a commonly used dataset for social network research, and is very well-suited to our setting: correspondents belong to a closed network of company employees resulting in a fairly dense network, and the text of emails is included in the dataset.  We considered all emails found in the Sent folders of ENRON-based email addresses, that were sent only to other ENRON-based email addresses, and excluded individuals who sent and received fewer than 10 emails. We removed standard stopwords, plus any words that occur more than 500 or fewer than 10 times in the corpus.  This resulted in a dataset with  a total of 48,064 non-stopwords sent between 90 email addresses, with a dictionary of length of 944.

\paragraph{Interactions in ``A Midsummer Night's Dream'':} Due to a lack of publicly-available email interaction networks, we supplement the ENRON dataset with an interaction network automatically generated from Shakespeare's ``A Midsummer Night's Dream''. We considered each speech a directed interaction from the speaker to the last person to speak; the first speech of each scene is not included in the dataset.

Admittedly, this dataset suffers limitations. The social network and interaction structure are not naturally occurring and are inherently stylized. Further, our data extraction method is imperfect: during multiple scenes between the Athenian characters, Puck and other fairy characters are on-stage but assumed invisible to the humans. Puck's asides and soliloquies are recorded as messages to the last human to speak, although this is not the author's intended interpretation. Despite these limitations, we find this dataset a useful addition since the main characters will be familiar to many readers, and naturally fall into a range of communities, such as the young Athenian lovers (Hermia, Lysander, Demetrius, and Helena) and the characters  in the play-within-a-play (Prologue, Lion, Pyramus, Thisbe, Wall, and Moonshine).

We removed standard stopwords, Elizabethan words that are equivalent to these stopwords, and the names of characters, plus words occurring more than 50 times in the play, resulting in a total of 5913 non-stopwords sent between 28 characters, with a dictionary of length 2,204.

\subsection{Comparison Methods}\label{sec:comparisons}

We compare the Topic Blockmodel against a range of comparison models, including models for text that take a network as a covariate; network models that ignore text; and standard topic models.

\begin{itemize}

\item \textbf{Latent Dirichlet allocation} (LDA) \citep{blei2003LDA}, a topic model that ignores network structure.

\item The \textbf{Author Recipient Topic Model} (ART) \citep{mccallum2005}, which uses the network as a covariate, and has a separate distribution over topics for each sender/recipient pair. 

\item A stochastic blockmodel with a gamma/Poisson link, which we will refer to as the \textbf{Poisson Stochastic Blockmodel} (Poisson-SBM). This can model the number of words exchanged, but not their content.

\item The \textbf{Clustered Node Topic Model} (CNT), a reduced version of the Topic Blockmodel which does not use a distribution over counts, instead conditioning on the observed counts.  This model begins with the same dsitribution over community assignments and, similar to the specification in Equation \ref{eqn:DirMultForm}, specifies a distribution for the vector of probabilities $\theta_{c_s,c_r}$ for each pair of communities, without any rate parameters.  In full,
\begin{equation}
  \begin{aligned}
    \phi \sim& \mbox{Dirichlet}_K(\xi_0)\\
    c_s \sim& \mbox{Dirichlet}(\phi),\; s=1,\dots,S\\
    \theta_{k,\ell} \sim& \mbox{Dirichlet}_T(\alpha_\lambda)\\
    z_{s,r,i} \sim& \mbox{Discrete}(\theta_{c_s,c_r}),\; i\in\{1,\dots, n_{s,r}^{(\cdot)}\}\\
    \eta_t \sim& \mbox{Dirichlet}_V(\kappa),\; t=1,\dots,T\\
    w_{s,r,i} \sim& \mbox{Discrete}(\eta_{z_{s,r,i}}).
  \end{aligned}\label{eqn:CNTModel}
\end{equation}

\end{itemize}

Due to the similarities between the models, all models were sampled using appropriately modified versions of the sampler described in Section \ref{sec:inference}. During the first 500 burn-in samples, we used simulated annealing to improve exploration, with the temperature set as $\tau = e^{1 - m/500}$, where  $m$ is the iteration. Hyperparameters were sampled with low-information priors using Metropolis-Hasting sampling. The number of topics was selected by cross validation, and the number of communities was set to $S/3$ for Shakespeare and $S/4$ for ENRON, where $S$ is the number of nodes. 

\subsection{Qualitative Evaluation}\label{sec:qual}

\begin{figure*}[p]
  \begin{center}
  \includegraphics[width=\textwidth]{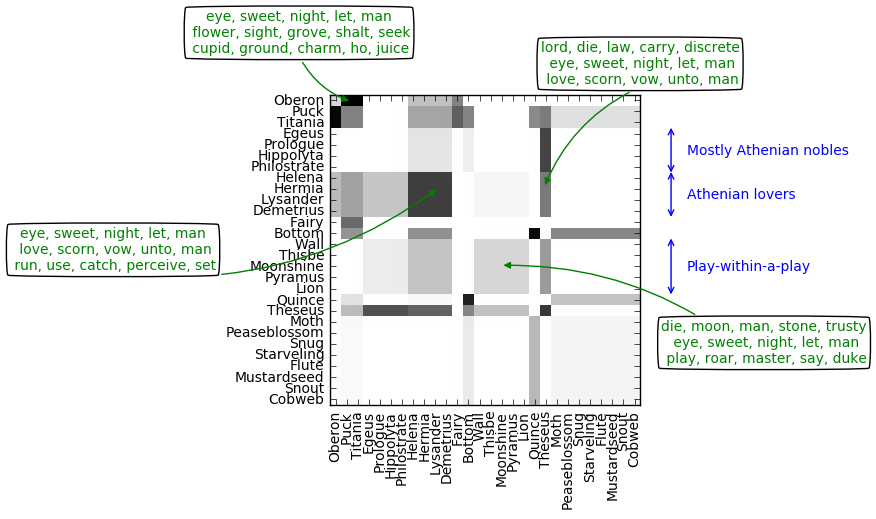}
  \caption{Communities found in ``A  Midsummer Night's Dream'', with highest-probability topics associated with community pairs.}
  \label{fig:shake_topics}
  \end{center}
\end{figure*}

\begin{figure*}[p]
  \begin{center}
  \includegraphics[width=\textwidth]{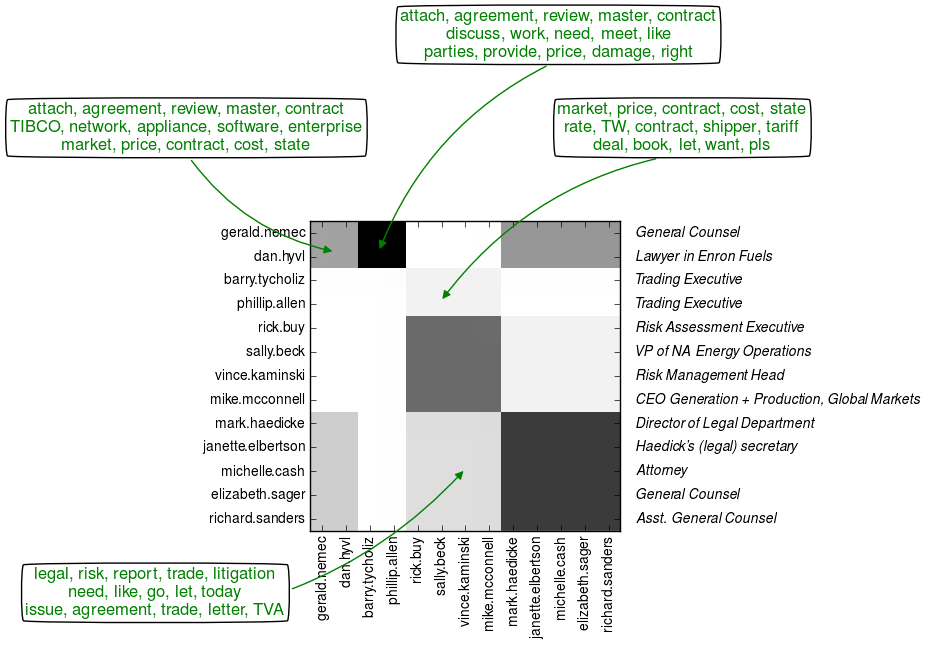}
  \caption{Communities found in the ENRON e-mail corpus for select e-mail participants, with highest-probability topics associated with community pairs.}
  \label{fig:enron_topics}
  \end{center}
\end{figure*}

\begin{figure*}[p]
  \begin{center}
  \includegraphics[width=\textwidth]{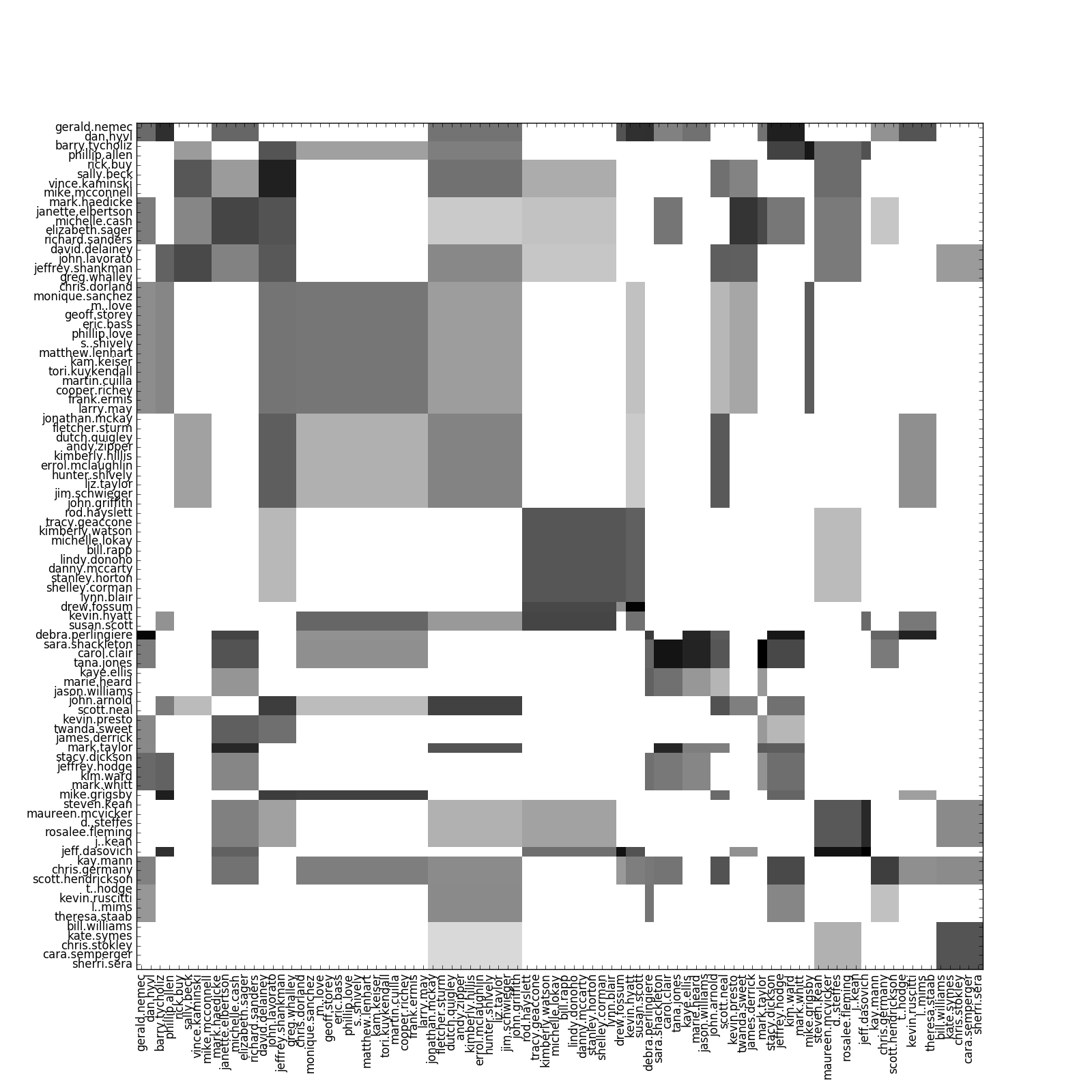}
  \caption{Communities found in the ENRON e-mail corpus for all ENRON internal e-mail participants.}
  \label{fig:enron_topics_full}
  \end{center}
\end{figure*}

We begin with a qualitative analysis of the community structure found using the Topic Blockmodel on ``A Midsummer Night's Dream'', since reader familiarity with the characters allow for easy evaluation of the clusterings found. Figure \ref{fig:shake_topics} shows the community structure obtained using a single sample from the Markov chain (to avoid alignment issues). Here, the shade of  element $(s,r)$ of the matrix represents the gamma random variable $\lambda_{c_s,c_r}^{(\cdot)}$ governing the total number of words sent from node $s$ to node $r$. The community structure can be inferred by looking at the discontinuities: nodes in the same community have the same parameter.

The names of the characters are given on the left hand axis, and some interesting communities are manually annotated on the right. Note that the communities generated are fairly well aligned with the character groupings present in the play.  For example, Demetrius, Helena, Hermia, and Lysander represent a ring of romantically entangled Athenians;  Egeus, Hippolyta and Philostrate are elder Athenian nobility; Wall, Prologue, Thisbe, Moonshine, Pyramus and Lion are all characters in the play-within-a-play; Titania and Puck are both fairies who interact with Oberon in a similar manner. The outliers are mostly characters with very few lines -- for example the minor fairies and the minor mechanicals are intermingled, but all these characters have very few lines.

To demonstrate how the topics characterize the community's relationships, we consider four community-community pairs that discuss love - a major theme of ``A Midsummer Night's Dream''. While all the selected pairs contain a shared topic of romantic words, the additional topics shed nature on the communities' nature. The star-crossed Athenian lovers talk among themselves of love and hate, and talk to Duke Theseus about the consequences of their romantic choices; Oberon talks to Puck and Titania of  magical slumber and fairy mischief; the play-within-a-play characters talk about aspects of the play and appeal to their audience.

Figures \ref{fig:enron_topics} and \ref{fig:enron_topics_full} show the discovered latent social network between a subset of the ENRON employees.  For ease of interpretability, Figure \ref{fig:enron_topics} provides an annotated subset of the Enron employees, their exchanged topics, and the employees' roles in the company; Figure \ref{fig:enron_topics_full} shows the full network. From Figure \ref{fig:enron_topics} we see that attorneys Gerald Nemec and Dan Hyvl are in a community, as are the trading executives Barry Tycholiz and Phillip Allen, as are executives involved in energy development and risk Rick Buy, Sally Beck, Vince Kamisnki, and Mike McConnell.  The fourth community shown is again legal professionals, but with different subject areas.

In the absence of this job title information, we could still use the topics associated with the community-community pairs to improve our understanding of the latent network. We see that  the attorneys' emails are focused on agreements and contracts, and supplying advice to the other employees. When the trading executives are talking with the development community, however, they are primarily discussing elements of economic forecasts (market, price, cost, rate, contract, tariff, etc.).  When the second attorney group is writing to the risk group, their topics skew more toward legal risks (e.g. litigation) and government affairs (e.g. dealing with the Tennessee Valley Authority (TVA)) than the contracts advice that the first group of attorneys gives to the trading executives.

\subsection{Quantitative Evaluation}\label{sec:quant}
We evaluated the predictive performance of our model on four metrics:
\begin{enumerate}
\item Log predictive likelihood of the text of held-out documents (conditioned on number of words sent, since this is required for most of the comparison methods). This is designed to mimic the task of predicting the topical content of an email from its sender and recipient.
\item Log predictive likelihood of the recipient of a held-out email/speech, conditioned on the sender and the text of the communication. This is designed to mimic the task of suggesting recipients for an email.
\item Log predictive likelihood of the sender and recipient of a held-out email/speech. This is designed to showcase the fact that using the text information allows us to better model latent community structure. 
\item Log predictive likelihood of the word counts of held-out sender-receiver pairs. This is designed to show that the inclusion of topic information improves count prediction.
\end{enumerate}

\begin{table}[ht]
	\caption{Log predictive likelihood ($\pm$ one standard error) of document text, conditioned on sender and recipient where applicable.}
	\centering
	\label{tab:topic_prediction}

	\begin{tabular}{l c c}
	Model & ENRON &  Shakespeare  \\
	\noalign{\smallskip}\hline\noalign{\smallskip}
	LDA	&				-410,110.2 	 $\pm$  50.8	&	--48,716.2	 $\pm$  4.6 \\
	ART &				-365,600.5 	$\pm$  47.7	&	-47,495.5  $\pm$  4.8 \\
	CNT	&			-368,983.5 	$\pm$  89.2	&	\textbf{-46,076.6} 	 $\mathbf{\pm}$  \textbf{3.9} \\
	\textit{Topic Blockmodel}	&				\textbf{-345.632.5} 	$\mathbf{\pm}$  \textbf{4.1}	&	-46,275.9 $\pm$  4.0 \\
	\noalign{\smallskip}\hline
	\end{tabular}
	
\end{table}

\begin{table}[h!]
	\caption{Log predictive likelihood ($\pm$ one standard error) of document recipient, conditioned on document content and sender where applicable.}
	\centering
        
	\label{tab:r_prediction}
	\begin{tabular}{l c c}
		Model & ENRON &  Shakespeare  \\ 
		\noalign{\smallskip}\hline\noalign{\smallskip}
		ART				& -204,585.3		$\pm$  6.4		&	-19,809.7 	$\pm$  1.1 \\
		CNT			& -216,278.9		$\pm$   $<$0.1		&	-19,703.3 	$\pm$  $<$0.1 \\
		 Poisson-SBM				& -160,984.7		$\pm$  148.6	&	-14,587.2	$\pm$  35.9 \\
		\textit{Topic Blockmodel}			& \textbf{-137,199.8} $\mathbf{\pm}$  \textbf{53.2}	&	\textbf{-12,997.8} 	 $\mathbf{\pm}$  \textbf{20.6} \\
	\noalign{\smallskip}\hline
	\end{tabular}
	
\end{table}

\begin{table}[ht]
	\caption{Log predictive likelihood ($\pm$ one standard error) of document sender and recipient, conditioned on document content where applicable.}
	\centering
        
	\label{tab:sr_prediction}
	\begin{tabular}{l c c}
		Model & ENRON &  Shakespeare   \\
		\noalign{\smallskip}\hline\noalign{\smallskip}
		ART				& -416,588.6		$\pm$  6.8		&	-39,580.0 	$\pm$  1.0 \\
		CNT				& -432,557.7		$\pm$   $<$0.1		&	-39,406.7 	$\pm$  $<$0.1 \\
		 Poisson-SBM			& -347,479.6		$\pm$  148.6	&	-31,400.3 	$\pm$  35.9 \\
		\textit{Topic Blockmodel}			& \textbf{-321,127.8} 	$\mathbf{\pm}$  \textbf{53.3}	&	\textbf{-29,614.0} 	 $\mathbf{\pm}$  \textbf{20.6} \\
	\noalign{\smallskip}\hline
	\end{tabular}
	
\end{table}

\begin{table}[ht]
	\caption{Log predictive likelihood ($\pm$ one standard error) of sender and recipient counts.}
	\centering
        
	\label{tab:count_prediction}
	\begin{tabular}{l c c}
		Model & ENRON &  Shakespeare   \\
		\noalign{\smallskip}\hline\noalign{\smallskip}
		 Poisson-SBM			& -92,851.2		$\pm$  12.1	&	-103,411.4 	$\pm$  0.6 \\
		\textit{Topic Blockmodel}			& \textbf{-88,730.4} 	$\mathbf{\pm}$  \textbf{3.1}	&	\textbf{-102,549.8} 	 $\mathbf{\pm}$  \textbf{0.2} \\
	\noalign{\smallskip}\hline
	\end{tabular}
	
\end{table}

\subsubsection{Log-likelihood of words in held-out documents}
For the first task, we randomly held out 10\% of documents, and evaluated the predictive log likelihood of this test set using the  comparison models with a topic model component (i.e.\ LDA, ART, and CNT). The log predictive likelihoods are shown in Table \ref{tab:topic_prediction}.

We see that the Topic Blockmodel performs significantly better than the competitors on the ENRON dataset. In this realistic setting, the number of emails sent between two individuals is highly indicative of their relationship, so we see a significant advantage from jointly modeling the number of words and their content. In particular, we see that the Topic Blockmodel outperforms our Clustered Node Topic Model variant, which does not model counts and treats zero edges as missing.

On the Shakespeare data, the Topic Blockmodel performs slightly worse than the Clustered-Node Topic model, though still better than LDA or ART. We believe that this is due to the artificial nature of the network. The community structure in ``A Midsummer Night's Dream'' is man-made, and designed so that the many separate communities interact in complex, artful manners. Moreover, by assuming a speech is directed to (only) the previous speaker, we are working with a noisy approximation to Shakespeare's intended interaction network.  Since the Clustered-Node Topic Model does not model the number of links, it will be less hampered by an unrealistic network structure. 

\subsubsection{Recipient Attribution}

For the second task, designed to mimic automatic email recipient suggestion, we again held out 10\% of documents and  predicted the recipient of each document based on the document's length, text and sender. 
We compared against the three comparison methods with a network component, namely ART, CNT, and the Poisson Stochastic Blockmodel. Prediction in ART and CNT does not take into account the number of words sent; prediction in the Poisson Stochastic Blockmodel does not take into account the specific words sent. Table \ref{tab:r_prediction} shows the test set log predictive likelihood for the four methods on the recipient attribution task.

In the ENRON e-mail data, we again see that the Topic Blockmodel performs significantly better than any of the competitive models in identifying the correct sender-recipient pair, with the Poisson Stochastic Blockmodel coming  second and the two models that do not consider word counts performing worst. The relative performance of the Poisson Stochastic Blockmodel (which does not consider topic distributions) versus CNT and ART (which do not consider word counts) suggests that count modeling, rather than topic modeling, is the more important component in this setting; however by combining these two components the Topic Blockmodel is able to make use of the topic distribution to improve prediction over the purely count-based model.

We see a similar pattern in the Shakespeare data: the Topic Blockmodel  outperforms the Poisson Stochastic Blockmodel and all other models, and the models that just consider topical content of documents perform worse than the Poisson Stochastic Blockmodel that only considers counts.  This is again likely for similar reasons to ENRON: the models on interaction intensity are able to down-weight pairs that very rarely interact, greatly boosting the likelihood of pairs that are expected to interact, and further identifying the correct topic mixture within high-intensity community pairings.

\subsubsection{Sender/Recipient Attribution}
For the third task, we again held out 10\% of documents and  predicted both sender and recipient based on a document's length and text, comparing against ART, CNT and the Poisson Stochastic Blockmodel. The resulting log predictive likelihoods, shown in Table \ref{tab:sr_prediction}, tell a similar story to the sender attribution task: the Poisson Stochastic Blockmodel, which only considers document length, outperforms CNT and ART which only consider document text, suggesting document length is more important than document semantic content in this task. However, the Topic Blockmodel, by making use of both length and semantic content, is able to outperform all three comparison methods on both tasks.

\subsubsection{Edge Count Prediction}
Finally, we withheld 10\% of sender-receiver pairs in the network and predicted the word count of the withheld links based on the assigned communities of the sender and receiver.  Table \ref{tab:count_prediction} shows that, in both the ENRON and Shakespeare data sets, the Topic Blockmodel significantly improves on the Poisson Stochastic Blockmodel, which is the only comparison model discussed which models the word counts of heldout links.

\section{Discussion and Future Work}\label{sec:conclusion}

In this paper we introduced a unified network and topic model, the Topic Blockmodel. Inspired by existing stand-alone network and topic models, the Topic Blockmodel can be used to identify and label communities in a network and make predictions about interactions. 

We have focused here on networks where the interactions are textual in nature. However, we may also have networks where interactions take the form of images, audio, or some combination of media. A future research direction might be to explore augmenting this model with other forms of media to better make use of information shared across the network, using likelihoods such as those described in \citep{cao2007spatially}, \citep{niu2012context} or \citep{kim2009acoustic}.

Other extensions could be obtained by using a richer distribution over the community structure. We chose a simple, parametric model with single-community membership to allow for straightforward computation; however the potential for mixed-membership or nonparametric versions is clear. Another interesting avenue for research is to make the distribution over communities explicitly dependent on some set of covariates such as time of email or geographical location of nodes, creating a dynamic model.

One limitation of the stochastic blockmodel framework is that it is only appropriate when our network is dense -- that is, when the number of non-zero edges grows quadratically with the number of nodes. This is a reasonable assumption in relatively small networks where it is likely that all nodes have had a chance to interact with each other -- for example, groups of individuals within a school, company or organization. This is the setting we have explored in this paper.

An interesting parallel line of research, which we are currently exploring, is models for text-based interaction in \textit{sparse} data. Such a model would require replacing the stochastic blockmodel component of the model with a distribution appropriate for sparse graphs, such as those described by \citep{Caron:Fox:2014}, \citep{Veitch:Roy:2015}, \citep{Cai:Campbell:Broderick:2016}, \citep{Crane:Dempsey:2016} and \citep{williamson2016nonparametric}. Without such a significant change to the model, one possible direction would be to add node-specific degree-correcting parameters as proposed by \citep{Karrer:Newman:2011}.


\bibliographystyle{plain}
\bibliography{network_topic}

\begin{thebibliography}{10}

\bibitem{acharya2015}
A.~Acharya, D.~Teffer, J.~Henderson, M.~Tyler, M.~Zhou, and J.~Ghosh.
\newblock Gamma process poisson factorization for joint modeling of network and
  documents.
\newblock In {\em Joint European Conference on Machine Learning and Knowledge
  Discovery in Databases}, 2015.

\bibitem{Airoldi:Blei:Fienberg:Xing:2008}
E.M. Airoldi, D.M. Blei, S.E. Fienberg, and E.P. Xing.
\newblock Mixed membership stochastic blockmodels.
\newblock {\em Journal of Machine Learning Research}, 9:1981--2014, 2008.

\bibitem{blei2006dynamic}
D.M. Blei and J.D. Lafferty.
\newblock Dynamic topic models.
\newblock In {\em International Conference on Machine Learning}, 2006.

\bibitem{ctm}
D.M. Blei and J.D. Lafferty.
\newblock A correlated topic model of {S}cience.
\newblock {\em The Annals of Applied Statistics}, 1(1):17--35, 2007.

\bibitem{blei2003LDA}
D.M. Blei, A.Y. Ng, and M.I. Jordan.
\newblock Latent {D}irichlet allocation.
\newblock {\em The Journal of Machine Learning Research}, 3:993--1022, 2003.

\bibitem{bouveyron2018stochastic}
Charles Bouveyron, Pierre Latouche, and Rawya Zreik.
\newblock The stochastic topic block model for the clustering of vertices in
  networks with textual edges.
\newblock {\em Statistics and Computing}, 28(1):11--31, 2018.

\bibitem{Cai:Campbell:Broderick:2016}
D.~Cai, T.~Campbell, and T.~Broderick.
\newblock Edge-exchangeable graphs and sparsity.
\newblock In {\em Advances in Neural Information Processing Systems}, 2016.

\bibitem{cao2007spatially}
L.~Cao and L.~Fei-Fei.
\newblock Spatially coherent latent topic model for concurrent segmentation and
  classification of objects and scenes.
\newblock In {\em International Conference on Computer Vision}, 2007.

\bibitem{Caron:Fox:2014}
F.~Caron and E.B. Fox.
\newblock Sparse graphs using exchangeable random measures.
\newblock arXiv:1401.1137 [stat.ME], 2017.

\bibitem{rtm}
J.~Chang and D.M. Blei.
\newblock Relational topic models for document networks.
\newblock In {\em International Conference on Artificial Intelligence and
  Statistics}, 2009.

\bibitem{Crane:Dempsey:2016}
H.~Crane and W.~Dempsey.
\newblock Edge exchangeable models for network data.
\newblock arXiv:1603.04571 [math.ST], 2016.

\bibitem{griffiths2004finding}
T.L. Griffiths and M.~Steyvers.
\newblock Finding scientific topics.
\newblock {\em Proceedings of the National Academy of Sciences}, 101(suppl
  1):5228--5235, 2004.

\bibitem{Karrer:Newman:2011}
B.~Karrer and M.E.J. Newman.
\newblock Stochastic blockmodels and community structure in networks.
\newblock {\em Physical Review E}, 83(1):016107, 2011.

\bibitem{irm}
C.~Kemp, J.B. Tenenbaum, T.L. Griffiths, T.~Yamada, and N.~Ueda.
\newblock Learning systems of concepts with an infinite relational model.
\newblock In {\em National Conference on Artificial Intelligence (AAAI)}, pages
  381--388, 2006.

\bibitem{kim2009acoustic}
S.~Kim, S.~Narayanan, and S.~Sundaram.
\newblock Acoustic topic model for audio information retrieval.
\newblock In {\em IEEE Workshop on Applications of Signal Processing to Audio
  and Acoustics}, 2009.

\bibitem{enron_paper}
J.~Leskovec, K.J. Lang, A.~Dasgupta, and M.W. Mahoney.
\newblock Community structure in large networks: Natural cluster sizes and the
  absence of large well-defined clusters.
\newblock {\em Internet Mathematics}, 6(1):29--123, 2009.

\bibitem{mccallum2005}
A.~McCallum, A.~Corrada-Emmanuel, and X.~Wang.
\newblock The author-recipient-topic model for topic and role discovery in
  social networks, with application to {E}nron and academic email.
\newblock In {\em Workshop on Link Analysis, Counterterrorism and Security},
  2005.

\bibitem{niu2012context}
Z.~Niu, G.~Hua, X.~Gao, and Q.~Tian.
\newblock Context aware topic model for scene recognition.
\newblock In {\em Computer Vision and Pattern Recognition}, pages 2743--2750,
  2012.

\bibitem{Snijders:Nowicki:1997}
T.A.B. Snijders and T.~Nowicki.
\newblock Estimation and prediction for stochastic blockmodels for graphs with
  latent block structure.
\newblock {\em Journal of Classification}, 14(1):75--100, 1997.

\bibitem{Teh:2007}
Y.~Teh, M.~Jordan, M.~Beal, and D.~Blei.
\newblock Hierarchical {D}irichlet processes.
\newblock {\em Journal of the American Statistical Association},
  101(476):1566--1581, 2007.

\bibitem{tu2010}
Y.~Tu, N.~Johri, D.~Roth, and J.~Hockenmaier.
\newblock Citation author topic model in expert search.
\newblock In {\em ACL International Conference on Computational Linguistics},
  2010.

\bibitem{Veitch:Roy:2015}
V.~Veitch and D.~Roy.
\newblock The class of random graphs arising from exchangeable random measures.
\newblock arXiv:1512.03099 [math.ST], 2015.

\bibitem{Wang:Wong:1987}
Y.J. Wang and G.Y. Wong.
\newblock Stochastic blockmodels for directed graphs.
\newblock {\em Journal of the American Statistical Association}, 82(397):8--19,
  1987.

\bibitem{williamson2016nonparametric}
S.A. Williamson.
\newblock Nonparametric network models for link prediction.
\newblock {\em Journal of Machine Learning Research}, 17(202):1--21, 2016.

\bibitem{Zhu:Yan:Getoor:Moore:2013}
Y.~Zhu, X.~Yan, L.~Getoor, and C.~Moore.
\newblock Scalable text and link analysis with mixed-topic link models.
\newblock In {\em ACM SIGKDD International Conference on Knowledge Discovery
  and Data Mining}, 2013.

\end{thebibliography}

\end{document}